\documentclass{vgtc}                          

\graphicspath{{figures/}{pictures/}{images/}{./}} 

\usepackage{times}                     

\usepackage{tabu}                      
\usepackage{booktabs}                  
\usepackage{lipsum}                    
\usepackage{mwe}                       
\usepackage[linesnumbered,ruled,vlined]{algorithm2e}
\usepackage{amsmath}

\usepackage{mathptmx}                  

\onlineid{1227}

\vgtccategory{Research}

\vgtcinsertpkg

\title{Fast Texture Transfer for XR Avatars via Barycentric UV Conversion}

\author{Hail Song\thanks{e-mail: hail96@kaist.ac.kr}\\
\scriptsize KAIST UVR Lab%
\and Seokhwan Yang\thanks{e-mail: ysshwan147@kaist.ac.kr}\\
\scriptsize KAIST UVR Lab%
\and Woontack Woo\thanks{e-mail: wwoo@kaist.ac.kr}\\ %
     \parbox{1.4in}{\scriptsize \centering KAIST UVR Lab \\ KAIST KI-ITC ARRC}}

\teaser{
  \centering
  \includegraphics[width=\linewidth]{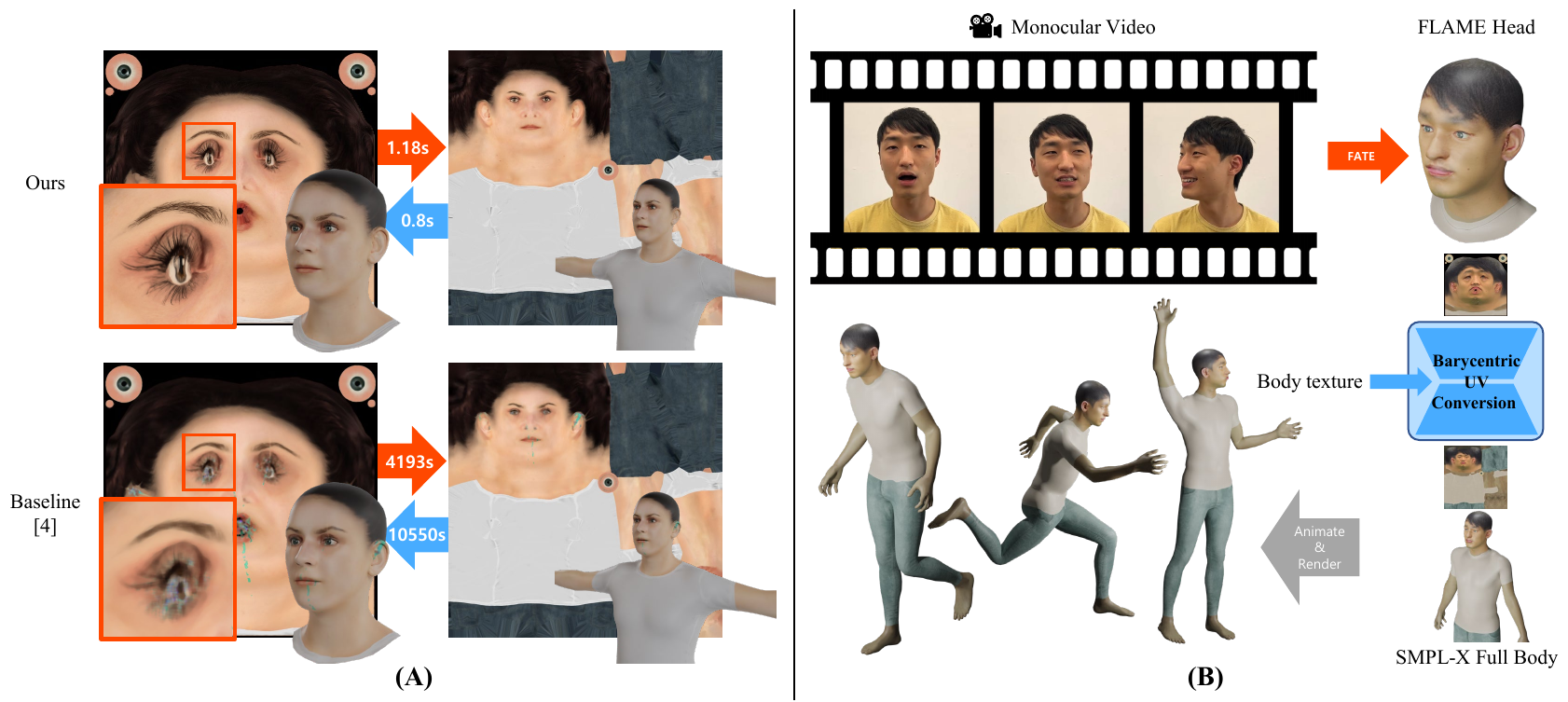}
  \caption{(A) Comparison of texture transfer results: baseline~\cite{CvHadesSun_Flame2SMPLX} vs. ours. (B) Application to selfie-based full-body avatar generation using FATE~\cite{zhang2025fate} and our method.}
  \label{fig:teaser}
}

\abstract{
    We present a fast and efficient method for transferring facial textures onto SMPL-X–based full-body avatars. Unlike conventional affine-transform methods that are slow and prone to visual artifacts, our method utilizes a barycentric UV conversion technique. Our approach precomputes the entire UV mapping into a single transformation matrix, enabling texture transfer in a single operation. This results in a speedup of over 7000x compared to the baseline, while also significantly improving the final texture quality by eliminating boundary artifacts. Through quantitative and qualitative evaluations, we demonstrate that our method offers a practical solution for personalization in immersive XR applications.
    The code is available online.\footnote{Project page: \url{https://github.com/hailsong/FastFLAME2SMPLX}}
} 


\begin{document}


\firstsection{Introduction}

\maketitle

Avatars in XR environments play a crucial role in conveying a user’s identity and enabling communication within virtual spaces. In particular, avatars that closely resemble the real user enhance both immersion and presence, thereby shaping the overall quality of the user experience. Recent studies aimed at faithfully reproducing a user’s appearance~\cite{zhang2025fate, song2023rc, song2024toward} build upon the parametric head model FLAME~\cite{FLAME:SiggraphAsia2017} and the full-body model SMPL-X~\cite{smplx}.
While existing research has successfully enabled realistic avatar representation for heads and full bodies individually, integrating these for detailed facial expression and full-body animation simultaneously faces limitations.
Specifically, these methods often entail lengthy processing times, which hinder their practical applicability.

In this work, we consider a scenario in which a selfie video captured by the user is used to extract facial textures and composite them onto an SMPL-X–based full-body avatar in near real time. The conventional per-face affine-transform approach~\cite{CvHadesSun_Flame2SMPLX} at the mesh-triangle level suffers from high computational cost, with processing time growing as the number of triangles increases.

To address this limitation, we introduce a barycentric UV conversion technique that consolidates the mapping from triangle coordinates to UV coordinates into a single matrix operation. Furthermore, to demonstrate the practical utility of texture transfer from a head model to a full-body model, we propose a pipeline that leverages a video-based head model reconstruction method~\cite{zhang2025fate} in conjunction with our approach. Through a round-trip conversion, where we transform SMPL-X texture to FLAME and then back to SMPL-X, we quantitatively compared image loss and processing time. Our method demonstrated over 7000 times faster texture conversion speed and achieved higher similarity to the original image compared to the baseline.

\section{Method}

Our method for transferring face textures consists of two main stages, as detailed in \autoref{alg:texture_transfer}. First, in a pre-computation stage, we build a complete sampling map that defines the geometric transformation between the source and target UV spaces. Second, in the transfer stage, we use this map to transform the texture in a single, high-speed operation.

\begin{algorithm}[htbp]
\caption{Fast Texture Transfer via Pre-computed Barycentric Map}
\label{alg:texture_transfer}
\KwIn{Source texture $T_{src}$, Target UV vertices $V_{tgt}$, Vertex correspondence map $M$}
\KwOut{Target texture $T_{tgt}$}
\BlankLine

\tcc{Stage 1: Pre-computation of a per-pixel sampling map ($S_{map}$)}
\ForEach{triangle $\triangle_{tgt}$ in $V_{tgt}$}{
    Find corresponding vertices for $\triangle_{tgt}$ from $M$ to form source triangle $\triangle_{src}$\;
    \ForEach{pixel $p_{tgt}$ within $\triangle_{tgt}$}{
        Calculate barycentric coordinates $(\alpha, \beta, \gamma)$ of $p_{tgt}$ with respect to $\triangle_{tgt}$'s vertices\;
        $p_{src} \leftarrow \alpha \cdot \triangle_{src}.\text{vertex}[0] + \beta \cdot \triangle_{src}.\text{vertex}[1] + \gamma \cdot \triangle_{src}.\text{vertex}[2]$\;
        \tcp{Store the source coordinate}
        $S_{map}(p_{tgt}) \leftarrow p_{src}$\;
    }
}
\BlankLine
\tcc{Stage 2: Texture Transfer}
$T_{tgt} \leftarrow H(T_{src})$ \\
\BlankLine
\Return{$T_{tgt}$}\;
\end{algorithm}

In the first stage (pre-computation), we establish the mapping using barycentric coordinates. The process begins with a pre-defined vertex correspondence map ($M$) that links the facial vertices between the source and target models. For any given pixel ($p_{tgt}$) in the target UV map, we first identify its containing triangle ($\triangle_{tgt}$) and compute its barycentric coordinates $(\alpha, \beta, \gamma)$. Using the map $M$, we find the corresponding triangle ($\triangle_{src}$) in the source UV map and calculate the precise source coordinate ($p_{src}$). This mapping from each target pixel to a source coordinate is stored in our sampling map, $S_{map}$.

In the second stage (texture transfer), this pre-computed map enables extremely high performance. The entire complex mapping process is consolidated, allowing the texture transfer to be executed as a single, efficient sampling operation. This is equivalent to a matrix multiplication across the whole image: $T_{tgt} = H(T_{src})$, where $H$ represents the transformation defined by $S_{map}$. This approach is significantly more efficient than conventional methods that perform individual affine transformations for each triangle. In our primary scenario, areas outside the face are seamlessly blended with a mean texture to ensure a natural appearance.

\section{Experiments and results}

We conducted experiments to evaluate our texture transfer method based on two key criteria: processing time to measure efficiency and image similarity to assess the quality of the converted textures.

Image similarity was measured using a round-trip conversion process. First, an original SMPL-X UV texture map was converted to a FLAME UV texture map using each algorithm. This FLAME texture was then immediately converted back into the SMPL-X UV space. The resulting texture was compared against the original SMPL-X texture to measure any degradation. During the FLAME-to-SMPL-X reverse conversion, the body part of the texture was populated with the original SMPL-X body texture, since the FLAME model only contains facial data.

The comparison was quantified using four standard image similarity metrics: L1 distance, Structural Similarity Index (SSIM), Peak Signal-to-Noise Ratio (PSNR), and Learned Perceptual Image Patch Similarity (LPIPS). For our evaluation, we used FLAME2SMPLX \cite{CvHadesSun_Flame2SMPLX} as the baseline method.
Both our method and the baseline approach were executed on a system equipped with a single RTX 3090 GPU, 128GB of RAM, and an Intel Core i9-10980XE CPU.
For the experiment, the SMPL-X and FLAME texture maps were processed at resolutions of 4096$\times4096$ and $2048\times2048$, respectively.


\begin{table}[!t]
\centering
\label{tab:results}
\begin{tabular}{@{}lccccc@{}}
\toprule
\textbf{Method} & \textbf{Time $\downarrow$} & \textbf{L1 $\downarrow$} & \textbf{SSIM $\uparrow$} & \textbf{PSNR $\uparrow$} & \textbf{LPIPS $\downarrow$} \\
\midrule
Baseline \cite{CvHadesSun_Flame2SMPLX} & 14743 s & 0.016 & 0.75 & 31.02 & 0.456 \\
Ours & \textbf{1.98 s} & \textbf{0.002} & \textbf{0.98} & \textbf{42.93} & \textbf{0.017} \\
\bottomrule
\end{tabular}
\caption{Performance comparison on the round-trip texture transfer task (SMPL-X $\to$ FLAME $\to$ SMPL-X). Our method significantly outperforms the baseline in both processing speed and texture quality.}
\vspace{-1\baselineskip}
\end{table}

The results of our experiments are summarized in \autoref{tab:results}. Our method demonstrates a dramatic improvement in both processing time and conversion quality. It achieves a speedup of over 7000x compared to the FLAME2SMPLX baseline while also significantly outperforming it across all image similarity metrics (L1, SSIM, PSNR, and LPIPS). The baseline's lower quality scores are further exacerbated by noticeable visual artifacts that appear along the facial boundary seams, an issue our method avoids. This indicates that our method not only runs faster but also maintains a much higher degree of fidelity to the original texture.
\autoref{fig:teaser} (A) qualitatively demonstrates that our method produces significantly higher-quality texture transfers, particularly around facial boundaries, compared to the baseline.
The one-time pre-computation stage for generating the UV grid takes approximately 9.77 seconds for SMPL-X to FLAME and 13.12 seconds for FLAME to SMPL-X. Once the UV grid is generated and cached, the texture transfer itself takes only 0.80 seconds for SMPL-X to FLAME and 1.18 seconds for the reverse direction, allowing for significantly faster processing during subsequent conversions.

Additionally, we present qualitative results to demonstrate a use-case scenario: generating a facial avatar from a selfie video and then creating a full-body avatar based on it. \autoref{fig:teaser} (B) shows the rendered result after acquiring a facial texture using FATE~\cite{zhang2025fate} and subsequently converting it to a full-body texture with our method. This demonstrates that by combining our method with existing head and full-body avatar generation pipelines, we can significantly accelerate the creation of personalized avatars that reflect a user's real appearance.


\acknowledgments{
This work was supported by the National Research Council of Science \& Technology (NST) grant by the Korea government (MSIT) (No. CRC21014) and the MSIT(Ministry of Science and ICT), Korea, under the Graduate School of Metaverse Convergence support program(IITP-2022(2025)-RS-2022-00156435) supervised by the IITP(Institute for Information \& Communications Technology Planning \& Evaluation)}

\bibliographystyle{abbrv-doi}

\bibliography{template}
\end{document}